
%
\input harvmac
\noblackbox
\def\Title#1#2{\rightline{#1}\ifx\answ\bigans\nopagenumbers\pageno0\vskip1in
\else\pageno1\vskip.8in\fi \centerline{\titlefont #2}\vskip .5in}

%
%
\def\ajou#1&#2(#3){\ \sl#1\bf#2\rm(19#3)}

\def\ie{{\it i.e.}}
\def\hf{{1\over2}}

\font\ticp=cmcsc10

\def\grad{\nabla}
\def\dee{\partial}
\def\act{{\cal L}}
%
%
\lref\str{A. Strominger,``Heterotic Solitons,''\ajou Nuc. Phys. &B343
(90) 167.}
\lref\klopp{R. Kallosh, A. linde, T. Ortin, A. Peet, and A. Van Proeyen,
 ``Supersymmetry as a cosmic censor,"\ajou Phys. Rev. &D46 (92) 5278.}
\lref\ort{Tomas Ortin,``Electric-magnetic duality and supersymmetry in
stringy black holes," \ajou Phys. Rev. &D47 (93) 3136.}
\lref\gps{S. B. Giddings, J. Polchinski, and A. Strominger,
  ``Four dimensional black holes in string theory," hep-th 9305083;
   to appear in Phys. Rev. D.}
\lref\cham{A. H. Chamseddine, ``$N=4$ supergravity coupled to $N=4$
 matter and hidden symmetries,"\ajou Nuc. Phys. &B185 (81) 403.}
\lref\witt{E. Witten, ``On string theory and black holes,"
 \ajou Phys. Rev. &D44 (91) 314.}
\lref\gibma{G. W. Gibbons and K. Maeda, ``Black holes and membranes in
 higher dimensional theories with dilaton fields," \ajou Nuc. Phys.
&B298 (88) 741.}
\lref\ghs{D. Garfinkle, G. Horowitz, and A. Strominger, ``Charged black
holes in string theory," \ajou Phys. Rev. &D43 (91) 3140;
erratum\ajou Phys. Rev. &D45 (92) 3888.}
\lref\gh{R. Gregory and J. A. Harvey, ``Black holes with a massive
dilaton," \ajou Phys. Rev. &D47 (93) 2411.}
\lref\chs{C. G. Callan, J. A. Harvey, and A. Strominger, ``Worldsheet
approach to heterotic solitons and instantons," \ajou Nuc. Phys.
&B359 (91) 611.}
\lref\gs{S. B. Giddings and A. Strominger, ``Exact black fivebranes in
critical superstring theory," \ajou Phys. Rev. Lett. &67 (91) 2930.}
\lref\masch{J. Maharana and J. H. Schwarz, ``Noncompact symmetries in
string theory,"\ajou Nuc. Phys. &B390 (93) 3.}
\lref\cahamast{C. G. Callan, J. A. Harvey, E. Martinec, and A.
Strominger, unpublished.}
\lref\sev{A. Sevrin, W. Troost, and A. van Proeyen, ``Superconformal
Algebras in Two Dimensions with N=4,"\ajou Phys.
Lett. &B208 (88) 447.}
\lref\gepwit{D. Gepner and E. Witten,``String Theory on Group
Manifolds,"\ajou Nuc. Phys. &B278 (86) 493.}
\lref\pdv{P. Di Vecchia, V.G. Knizhnik, J.L. Peterson, and P. Rossi,
``A supersymmetric Wess-Zumino lagrangian in two dimensions,''
\ajou Nuc. Phys. &B253 (85) 701.}
\lref\kasu{Y. Kazama and H. Suzuki,\ajou Nuc. Phys. &B321 (89) 232.}
\lref\powi{A. Polyakov and P. Wiegmann,``Theory of nonabelian goldstone
bosons in two dimensions,"\ajou Phys. Lett. &B131 (83) 121.}
\lref\bars{I. Bars and K. Sfetsos,``A superstring theory in four curved
spacetime dimensions,"\ajou Phys. Lett. &B277 (92) 269.}
\lref\braden{H. Braden,``Paralellizing torsion and anomalies,''\ajou
 Phys. Rev. &D33 (86) 2411.}
\lref\calth{C. Callan and L. Thorlacius,``Sigma models and string
theory,'' in Particles, Strings and Supernovae, ed. A. Jevicki and
C-I. Tan (lectures from TASI 1988).}
\lref\banks{T. Banks and L. Dixon,``Constraints on string vacua with
spacetime supersymmetry,''\ajou Nuc. Phys. &B307 (88) 93.}

\Title{\vbox{\baselineskip12pt\hbox{UCSBTH-93-10}
\hbox{hepth@xxx/9312058}
}}
{\vbox{\centerline {Kaluza-Klein Black Holes}\vskip2pt
\centerline{ in String Theory}
}}

\centerline{{\ticp William  Nelson}\footnote{$^\dagger$}
{Email address: nelson@denali.physics.ucsb.edu}}
\vskip.1in
\centerline{\sl Department of Physics}
\centerline{\sl University of California}
\centerline{\sl Santa Barbara, CA 93106-9530}
\bigskip
\centerline{\bf Abstract}
Exact solutions of  heterotic string theory corresponding to four-dimensional
magnetic black holes in $N=4$ supergravity are described. The solutions
describe the black holes in the throat limit, and consist of a tensor
product of an $SU(2)$ WZW orbifold with the linear dilaton vacuum,
 supersymmetrized to $(1,0)$ world sheet SUSY. One dimension of
the $SU(2)$ model is interpreted as a compactified fifth dimension,
leading to a four dimensional solution with a Kaluza-Klein gauge field
having a magnetic monopole background; this corresponds to a solution in
$N=4$ supergravity, since that theory is obtained by dimensional
reduction of string theory.

\Date{9/93}

\newsec{Introduction}

The study of black holes in string theory has been pursued quite
vigorously in the last few years,
 with the hope that long-standing puzzles
may be resolved in the context of an apparently consistent
(but incomplete) theory of
quantum gravity.

 The first place to search for black hole solutions
is in the low energy effective theory of string theory in four
dimensions, and many have been found. They fall into two categories:
those that involve only the fundamental fields of string theory,
namely the metric, antisymmetric tensor, and $E_8\times E_8$ gauge fields;
and those that involve other fields resulting from compactification, such
as Kaluza-Klein gauge fields and scalars, etc. Solutions in the first category
include the four dimensional solutions of refs.
\refs{\gibma, \ghs}, as well as the five-dimensional solutions of
\refs{\str, \chs}. Into the latter category fall the black
holes of $N=4$ supergravity \refs{\klopp },
since all the gauge fields of that
theory arise from reduction of the ten dimensional metric and
antisymmetric tensor \refs{\cham}.
 These are the solutions of interest in this paper.

After studying solutions of the low energy effective action, it is
desirable to continue and find the corresponding exact vacua for string
theory, i.e. backgrounds whose sigma model is conformally invariant.
The first such exact black hole was constructed in \refs{\witt};
this was a two
dimensional solution. Five dimensional solutions were studied in
\refs{\chs , \gs}. Here two types of exact solution were found:
asymptotically flat solutions having sufficient supersymmetry to be
unrenormalized, and solutions corresponding to the infinite throat of an
extremal black  hole, without the asymptotically flat region.

The first four dimensional exact solutions, corresponding to the
magnetic black holes of \refs{ \gibma ,\ghs}, were constructed
in \refs{\gps}; here only the infinite throat of the extremal
black hole was exactly constructable. In this paper we show that
a similar construction
 gives the throat limit of some black holes in the
second category above, namely extreme magnetic black holes in $N=4$
supergravity.

The outline of the paper is as follows. In section two we discuss the low
energy
solution whose exact analog we will find. We give it as a solution of
five dimensional string theory, with compact fifth dimension,
 and show the correspondence with $N=4$
theory.  In section three we give the corresponding
exact solution in bosonic string theory, and in section four we discuss
supersymmetrization of the model to a heterotic model.
Finally, in section five we  explore the low-lying levels of the theory,
and construct operators giving throat-widening deformations of the
model.

\newsec{Low Energy Solution}

 $N=4$ supergravity in four dimensions is a truncation of
dimensionally reduced ten dimensional $N=1$ supergravity, which in turn
is the low energy theory of the heterotic string, with the $E_{8}\times
E_{8}$ degrees of freedom truncated. (Note that by truncation we mean
removal of fields from a theory in a way consistent with the equations
of motion.) Therefore to find
the exact string solution corresponding to an $N=4$ solution, one must
work in dimensions greater than four and find the appropriate
correspondence between fields. The solution we are interested in here is
the extreme magnetic black hole; in particular, only one of the six
$N=4$ gauge fields is nonvanishing, suggesting that we  need
only one extra dimension to find the string analog. Thus we begin with
the bosonic action for string theory in five dimensions,
with $E_8\times E_8$ gauge fields truncated, and assuming an unspecified
internal configuration for the remaining five dimensions:
$$S_{5}=\int d^{5}x{\sqrt{-\hat g}}e^{-2\hat\phi}\left[{\hat
  R}+4(\grad\hat\phi)^{2}-{1\over 12}{\hat H}^2 \right]
$$
where hats refer to five-dimensional objects,
and ${\hat H}_{\mu\nu\rho}\equiv ({\hat B}_{\mu\nu ,\rho}+
{\hat B}_{\nu\rho ,\mu}
     +{\hat B}_{\rho\mu ,\nu}).$
We will assume the fifth co-ordinate $x^5$ has a range $0\to 4\pi$; this
is its natural range in later sections, and it enters into the
dimensional reduction as a contribution to the dilaton shift (see
below).
Now we reduce to four dimensions, defining the four
dimensional  fields $g_{\mu\nu}, \ell, A^1, A^2, \phi ,B_{\mu\nu}$
by
\eqn\rdctn{ {\hat g}_{\mu\nu}=
\left(\matrix{g_{\mu\nu}+\ell^2 A_{\mu}^{1} A_{\nu}^1 &\ell^2 A_{\mu}^{1} \cr
   \ell^2 A_{\nu}^{1} &\ell^2 \cr}\right)
}
$$\eqalign{
  A^{2}_{\mu}&={\hat B}_{\mu 5}\cr
  \phi&={\hat\phi}-\hf\ln 4\pi\ell\cr
  B_{\mu\nu}&={\hat B}_{\mu\nu}.\cr}$$
With these definitions, standard formulas give (see e.g. \refs{\masch})
$$S_{4}=\int d^{4}x\sqrt{-g}e^{-2\phi}\left[ R+4(\grad\phi )^2
- {{\grad\ell}^2\over \ell^2}
-{\ell^2 \over 4}(F^{1})^2 +{1\over 4\ell^2}(F^{2})^2
  -{1\over 12}(H+A^1\wedge F^2)^2 \right]
$$
The term involving $H$ may be dualized into a
scalar axion, leading to its replacement by
$\hf(\grad a)^2 +{1\over 4}(F^{1}{\tilde F^{2}})$. Now we truncate the
theory, following \refs{\cham},
 by noticing that $\ell=\ell_{0}={\rm const.}$ is consistent with
the equations of motion if ${1\over\ell_{0}}F^{2}=\pm \ell_{0} F^{1}$. With
this
motivation we define new gauge fields
$$A^{\pm}={1\over \sqrt{2}}({\ell\over 2}A^{1}\pm{1\over 2\ell}A^{2})$$
and one finds that $\ell=\ell_0$ , $A^-=0$ is a consistent truncation.
Defining finally $F\equiv F^{+}$, the action becomes
$$I=\int d^{4}x\sqrt{-g}e^{-2\phi}\left[ R+4(\grad\phi)^2 -F^2+\hf(\grad a)^2
    +{\hf}aF{\tilde F}\right] .$$
This is the bosonic action of $N=4$ supergravity, with $5$ of $6$ vector
fields removed. We will be considering only purely magnetic black hole
solutions, which satisfy $F{\tilde F}=0$; thus we will take $a=\rm
   const.$ These black holes are a subset of those described in
\refs{\klopp}.

Now we write down the low energy extremal magnetic
black hole solution, using the
five dimensional variables.  With co-ordinates $t,r,\theta,\phi,\zeta$
the metric is
\eqn\loemet{ds^2=-dt^2+dr^2+R^{2}d\theta^2
+(R^{2}\sin^2\theta +Q^{2}\ell_{0}^{2}\cos^{2}\theta )d\phi^2
   +\ell_{0}^{2}d\zeta^2+2\ell_{0}^{2}Q\cos\theta d\phi d\zeta}
where $Q$ is a parameter measuring charge, and $R=R(r)$ satisfies
$$R^{\prime}=1-{Q\ell_{0}\over R}.$$
Note $r\in (-\infty ,\infty )$.
The antisymmetric tensor and dilaton are
\eqn\loerst{\eqalign{{\hat B}_{\phi\zeta}&=\ell_{0}^{2}Q\cos\theta\cr
  {\hat\phi}-{\hat\phi}_0 &=-\hf\ln\left( 1-{Q\ell_0 \over R}\right) .\cr}}
The dimensionally reduced quantities are determined from the above
using \rdctn.
    In particular one finds $A^{-}=0$, so this
is a solution of the truncated theory as well; therefore it corresponds
to a solution of the $N=4$ theory. The solution describes a
spherically symmetric asymtotically flat black hole, with no singularity
but rather an infinitely long throat having asymptotic radius
$Q\ell_0$.

Before going on to consider the corresponding exact solution, we pause
to
discuss the supersymmetry properties of this background. In
\refs{\klopp}, it is shown that this solution keeps two supersymmetries
of the $N=4$ theory unbroken; thus the five dimensional solution, with
the other five dimensions flat, should preserve half the supersymmetries of
the $D=10$, $N=1$ theory; and in fact this may be checked. One finds
that unbroken supersymmetries are generated by four spinors
$\epsilon_a$ having definite four-dimensional chirality (in the four
dimensions spanned by $r,\theta,\phi,\zeta$).
Furthermore,  the spinors $\epsilon_a$ may be
used to construct three complex structures, covariantly constant with
respect to the generalized connection
$\Omega^{-}\equiv {\hat\omega}-\hf{\hat H}$;
 this leads to enhanced $(4,0)$ worldsheet
supersymmetry for the heterotic sigma model with this
background (see also \refs{\chs}, and references therein, for similar
constructions and studies of extended supersymmetry in sigma models.)

Lastly we discuss the topology of the five dimensional solution.
For the solution \loemet, one might expect the topology
$R^2\times S^2\times S^1$; however, the monopole nature of the fields
makes this impossible, and requires
$S^1$ to be fibred over  $S^2$
 in a non-trivial way. What will appear in the sequel (and
what appeared in \refs{\gps}) is the space $R^2\times S^{3}$,
where $R^2$ is parametrized by $t,r$ and $S^3$ by $\theta , \phi ,
\zeta$ (and the $S^3$ may have further discrete identifications
due to orbifolding.)

\newsec{The Exact Solution}

We are looking for a conformally invariant sigma model to describe the
black hole throat. In this limit, the function $R$ is constant, and the
radial direction decouples from the angular directions and becomes flat
(but the dilaton will still vary with $r$). Thus to begin with we look for a
conformally invariant sigma model for the angular directions alone. This
is provided by an $SU(2)$ WZW model, as follows. For $g\in\rm SU(2)$,
we use the Euler angle parametrization
$$g=e^{i\phi S_3}e^{i\theta S_2}e^{i\left(\zeta-\phi\right) S_3}$$
where $S_{i}=\hf\sigma_{i}$, and
$\theta\in [0,\pi]$, $\phi\in [0,2\pi]$, $\zeta\in [0,4\pi]$.
Then for the two components of the WZW action
at level $k$ (bosonic for now) we
compute
\eqn\act{\eqalign{S_1&={k\over 16\pi}\int d^2 x{\rm Tr}
  \dee_\alpha g^{-1}\dee^\alpha g\cr
 &={k\over 16\pi}\int d^2 x\left[\hf(\grad\theta)^2
+(1-\cos\theta)(\grad\phi)^2 +\hf(\grad\zeta)^2
-(1-\cos\theta)\grad_\alpha \phi \grad^\alpha \zeta \right] \cr
S_2 &={k\over 24\pi}\int d^3 x\epsilon^{\alpha\beta\gamma}
 {\rm Tr}\left[ g^{-1}\dee_\alpha gg^{-1}\dee_\beta gg^{-1}\dee_\gamma  g
  \right] \cr
 &=-{k\over 16\pi}\int d^3 x\epsilon^{\alpha\beta\gamma}\sin\theta
  \dee_\alpha \theta\dee_\beta \phi\dee_\gamma \zeta .\cr} }
Comparing this to the sigma model action (with conveniently chosen
$\alpha^\prime$)
$$S={1\over 32\pi}\int d^2 x\dee_\alpha X^\mu \dee_\beta X^\nu
 \left(\eta^{\alpha\beta}G_{\mu\nu}
+\epsilon^{\alpha\beta}B_{\mu\nu}\right) $$
gives the metric and antisymmetric tensor. The metric comes out to be
$$G_{\mu\nu}=\left( \matrix{k&0&0\cr
 0 &k\sin^{2} \theta +k(\cos\theta-1)^{2} & k(\cos\theta-1) \cr
 0 & k(\cos\theta-1) & k \cr}\right)
$$
with the co-ordinates ordered $\theta,\phi,\zeta$, and the torsion is
(up to gauge)
$${\hat B}_{\phi\zeta}=k(\cos\theta-1) .$$
We see this takes the form of the solution \loemet; note that
$A^1_\phi=\cos\theta-1$ is a gauge transform of
 $A^1_\phi=\cos\theta$, with gauge parameter $\Lambda=\phi$. The parameters
have the values
$$\eqalign{Q&=1\cr \ell_{0}^2 &= k.\cr}$$

Thus we have an interpretation of the $SU(2)$ WZW model as an exact
bosonic string vacuum corresponding
 to a black hole in $N=4$ supergravity
(in the throat limit.)
To get the complete black hole throat model in heterotic string theory,
 we must add the  radial
theory, which is the linear dilaton vacuum, and we must supersymmetrize
both to $(1,0)$ supersymmetry. Also we must add an internal conformal
field theory to fill out a central charge of $(15,26)$

Lastly, we note the possibility of extending these models by
orbifolding; this is accomplished by  identifying points on the
$SU(2)$ group manifold  under $\zeta\to\zeta+{4\pi\over m}$.
 Modular invariance requires $m$ to divide $k$, thus
$k=mn$  (see \refs{\gps}).
To maintain the same co-ordinate range for the fifth co-ordinate, we
define $\chi$ by $\zeta={\chi\over m}$; then $\chi\in [0,4\pi]$
(recall that this range was assumed in \rdctn ).
 Then $G_{\mu\nu}$ may be rederived,
and one finds the parameter values
$$\eqalign{Q&=m\cr \ell_0^2&={n\over m}.\cr}$$
Thus one has a two-parameter family of models, with angular theory
$${SU(2)_{mn}\over Z(m)_R}.$$ However, in the remainder of the paper we
focus on the case $m=1$, since $m>1$ complicates the expressions without
any qualitative change.

\newsec{The Heterotic Sigma Model}

In this section we discuss the $(1,0)$ supersymmetric
extension of the the WZW action
\act\  .
To supersymmetrize we start by coordinatizing the group manifold, so
$g=g(x)$, $x\equiv (x^1,x^2,x^3)$.
We will also need a basis for tangent space;
 for this a good choice is the
left-invariant  vierbein
\eqn\lft{{L^a}_\mu={\sqrt 2}i\,{\rm Tr}[S^a g^{-1}\dee_\mu g ],}
where the normalization is such that ${L^a}_\mu {L^a}_\nu$ is the group
metric $g_{\mu\nu}=-{\rm Tr}[g^{-1}\dee_\mu gg^{-1}\dee_\nu g]$.
The meaning of
``left invariant'' is  the following: given a fixed element $h\in G$,
we define a mapping $G\to G$ by left multiplication, $g\to hg$; or, in
terms of coordinates, $x\to x^{\prime}(x)$, where $x^{\prime}$
has the appropriate form. A vierbein is
pushed forward by this mapping:
\eqn\viermap{e_\mu^a\to e_\mu^{\prime a}(x^\prime)\equiv
{\dee x^\nu \over \dee x^{\prime\mu}}e_\nu^a(x) .}
Then the vierbein is defined to be left-invariant if it is invariant
under this mapping :
\eqn\lftinv{e_\mu^{\prime a}(x^\prime) = e_\mu^a(x^\prime).}
One can verify that ${L^a}_\mu$ has this property.
Left invariance is defined similarly for other types of tensor field.

Now we write the $(1,0)$ sigma model action, using the $(1,0)$
superfield
\eqn\sf{X^\mu=x^\mu+\theta {E^\mu}_a(x) \psi^a,}
where ${E^\mu}_a$ is the inverse to ${L^a}_\mu$. The action is
\refs{\calth}
\eqn\sact{\eqalign{S_1&={1\over 32\pi}\int d^2xd\theta G_{\mu\nu}(X)
  D_+X^\mu\dee_-X^\nu \cr
 S_2&={1\over 32\pi}\int d^2xd\theta B_{\mu\nu}(X)
  D_+X^\mu\dee_-X^\nu ,\cr}}
where $D_+\equiv {\dee\over d\theta}+i\theta\dee_+$ is the
supercovariant derivative, and $G$,$B$ are the metric and torsion of the
original bosonic WZW model \act\ . Working out this action in
components, one finds \refs{\braden}
\eqn\compact{S_1+S_2={\rm WZW}_k+{k\over 8\pi}\int d^2x \psi_a\dee_+\psi^a.}
It is just the bosonic WZW model at level $k$, plus free fermions.

This is simple enough, but confusion can arise when one considers
different ways of introducing the fermions. In particular, choosing a
different
vierbein for the fermion part of equation \sf\ leads in general to a
different theory; tangent space rotations are anomalous here.
Furthermore, we now argue that in general one
gets a theory in which the Kac-Moody symmetries $g\to f(z)gh(\bar z)$
are anomalous; this is a disaster since without these symmetries there
is no reason to expect conformal invariance for the model.
The argument is simple: start with the
action \compact, and assume a different basis for the fermions, so the
fermion part becomes
\eqn\anferm{\int d^2x\psi_a{(M^{-1})^a}_b\dee_+({M^b}_c\psi^c),}
where ${M^a}_b(x)$ is some $x$-dependent rotation in tangent space;
the inverse vierbein ${E^\mu}_a{M^a}_b$ will be left-invariant if and only if
${M^a}_b$ is constant.
 Now, consider the transformation $x\to
x^{\prime}$ corresponding to $g\to f(z)gh({\bar z})$. This leaves the
${\rm WZW}_k$ part of the action invariant, as usual,
but on the fermions one
requires the extra
  rotation  $\psi^a\to
{{M^{-1}}^a}_b(x^{\prime}){{M^b}_c(x)}\psi^c$.
But such a rotation is anomalous; therefore the Kac-Moody symmetry is as
well.
Thus, to get the desired model, we must include the fermions with a
left-invariant vierbein, as in \sf; in this case the fermions are
free.

\newsec{The Spectrum}

The previous section discussed the $(1,0)$ supersymmetric WZW model from
the point of view of the heterotic sigma model. The
end result was that the model contains a bosonic WZW model at some level
$k$ plus three free fermions. In this section we discuss the $(1,0)$
model further, beginning from this  point.
We also add fields corresponding to the radial direction, which make up
the $(1,0)$  linear dilaton vacuum.
 We explore the low lying levels of the theory, and
construct operators corresponding to throat-widening deformations.

The content of the model is as follows: right and left moving level
$k$ $SU(2)$ Kac-Moody algebras,
 denoted $J^i(z)$, ${\tilde J}^i({\bar z})$; a free
boson (the radial coordinate), denoted $r$; and finally
 four free right moving fermions $\psi^a$ for $(1,0)$ supersymmetry.
 The fermions are treated exactly as in the
usual $D=10$ heterotic
string, \ie\ both R and NS sectors are included with the
standard GSO projection \refs{\bars}.
This, along with modular invariance of the
bosonic model \refs{\gepwit}, guarantees modular invariance of the full
theory. It also guarantees spacetime supersymmetry and absence of
tachyons.

As discussed in section $2$, the black holes in question have $N=2$
spacetime supersymmetry; this implies that they must have
 $(4,0)$ supersymmetry on the world sheet \refs{\banks}.
Furthermore, the $SU(2)$ Kac-Moody algebra which is contained in the
$(4,0)$ supersymmetry algebra must have level $1$, and therefore
 cannot be the same as the $SU(2)_k$ of
the bosonic theory. Thus this model should contain quite a large
symmetry algebra, a fact which has been noted and studied before in
 \refs{\sev,\chs, \cahamast}.
 The right-moving $(4,0)$
supersymmetry algebra consists of the stress tensor $T$, four
supercharges $G^{i}$, and an $SU(2)$ current algebra at level
one, given as follows:
\eqn\supalg{\eqalign{T&=-\hf\dee r\dee r-{1\over k+2}J^{i}J^{i}
   -\dee\psi^{a}\psi^{a}-{1\over \sqrt{2k+4}}\dee^2r  \cr
    G^{0}&=\sqrt{2}\dee r\psi^{0}+{2\over\sqrt{ k+2}}J^{i}\psi^{i}
           +{4\over\sqrt {k+2}}\psi^{1}\psi^{2}\psi^{3}
            -{2\over\sqrt{ k+2}}\dee\psi^0 \cr
    G^{1}&=\sqrt{2}\dee r\psi^1 +{2\over\sqrt {k+2}}\left(-J^1\psi^0+J^2\psi^3
            -J^3\psi^2\right) -{4\over\sqrt {k+2}}\psi^0\psi^2\psi^3
            -{2\over \sqrt{k+2}}\dee\psi^1\cr
    G^2 , G^3 &= {\rm cyclic \,\, perms.} \cr
    A_{i}^{-}&=\psi^0\psi^i +\hf\epsilon_{ijk}\psi^j\psi^k. \cr}}
The derivative terms at the end of the expressions for $T$ and $G^a$
are due to the linear dilaton in the $r$ direction. This algebra closes
on itself and has $c=6$
\refs{\cahamast}. Also on the right moving side another level $1$
$SU(2)$ algebra may be constructed from the fermions, namely
\eqn\kmalg{A_{i}^{+}=-\psi^0\psi^i +\hf\epsilon_{ijk}\psi^j\psi^k .}
Meanwhile the left moving side has only the current algebra ${\tilde J}^i$
 at level $k$, with
 stress tensor as above, minus fermion terms. Left
moving central charge is $4$.

Next we construct an operator corresponding to a throat widening
deformation, as evidence that the model can be connected to an
asymptotically flat region.
Such a deformation should alter $R(r)$ from constant to increasing; at
the same time, it should keep $\ell$ constant, to maintain the
correspondence with the $N=4$ theory.
 For the purely bosonic theory such an operator was
described in \refs{\gps}; it is
\eqn\bosvert{V_b=({\tilde J}^i {\tilde\phi}^1_i )(
 J^1\phi^1_1 +J^2\phi^1_2)e^{\alpha r},}
where $\phi^j_m$,${\tilde\phi}^j_{\tilde m}$ are the WZW primary fields,
and $\alpha$ will be determined later.
For the heterotic model we must supersymmetrize the right moving part,
i.e. find an operator with the same bosonic content which is the highest
component of a superfield. To do this
we start with the superconformal primary field (lowest component of a
superfield)
\eqn\scfprm{W={-\sqrt{k+2}}
\left(\psi^1\phi^1_1+\psi^2\phi^1_2\right)e^{\alpha r}.}
Using the operator product
\eqn\opprod{J^{i}(z)\phi^1_j(0) \sim {1\over z}\epsilon^{ijk}\phi^1_k}
one sees that $W$ has  a $1\over z$ singularity in its operator
product with $G^0$, as it should.
 The coefficient of $1\over z$ is the
 supersymmetric
operator we are looking for. It is
\eqn\ver{V=\left(J^1\phi^1_1+J^2\phi^1_2-4\phi^1_3\psi^1\psi^2\right)
 e^{\alpha r}-\alpha{\sqrt{2k+4}}
\psi_0\left(\psi^1\phi^1_1+\psi^2\phi^1_2\right)
 e^{\alpha r}}
where $\alpha=\sqrt{2\over k+2}$  to give dimension $(1,1)$.
Note that this   deformation
 breaks the
supersymmetries asociated to $G^i$, leaving only $(1,0)$, and in
particular eliminating spacetime supersymmetry. Thus the deformation is
away from extremality.

Next we discuss the low lying spectrum of the model, checking for
massless or tachyonic states. The vertex
operators (evaluated at $z={\bar z}=0$) take the general form
\eqn\vertop{e^{-\beta r}e^{ipr}({\tilde j}_{\tilde K}
 \cdot {\tilde\phi}^j_{\tilde m})
  (j_{K}\cdot {\phi}^j_m) V_{\rm ferm}V_{\rm int}
   {\tilde V}_{\rm int} }
where $j_K$ and ${\tilde j}_{\tilde K}$ represent  products of
WZW raising operators, $V_{\rm ferm}$ comes from the right moving
fermions $\psi^a$, and the operators labelled ``int" derive from the
internal theory.
Ghost factors have not been written in this expression.
The term $e^{-\beta r}e^{ipr}$ is a primary field of
the linear dilaton theory of the $r$ direction, where  $\beta$ must be chosen
such that the operator has real conformal weight for all real values of
$p$. Noting the coefficient of the $\dee^2r $ term in $T$, this fixes
$\beta={1\over \sqrt{2k+4}}$, giving the operator a weight of
$$L_0 = {\tilde L}_0=\hf p^2 +\hf{1\over 2k+4}.$$

Now we study the spectrum, beginning with the Ramond sector. Working
in the $-\hf$ picture, a state must have weights $(0,1)$, thus
\eqn\ram{\eqalign{0&=\hf p^2 +\hf{1\over 2k+4}+
    {j(j+1)\over k+2}+\vert K\vert+\hf N_f+L_0^{\rm
          int}\cr
 1&=\hf p^2+\hf{1\over 2k+4}
+{j(j+1)\over k+2}+\vert{\tilde K}\vert +{\tilde L}_0^{\rm int}.\cr}}
{}From the first equation above, we immediately have
$$p^2 < -{1\over k+2}\,,$$ thus ruling out massless and tachyonic states
in the R sector. In the NS sector, we work in the $-1$ picture, in which
states have weights $(\hf,1 )$, giving
\eqn\ns{\eqalign{\hf &=\hf p^2 +\hf{1\over 2k+4}+
{j(j+1)\over k+2}+\vert K\vert +\hf N_f
    +L_0^{\rm int}\cr
  1&=\hf p^2+\hf{1\over 2k+4}+
{j(j+1)\over k+2}+\vert{\tilde K}\vert+{\tilde L}_0^{\rm
     int}.\cr}}
Evidently $p^2 \leq 0$ requires $\vert K\vert =\vert {\tilde K}\vert =N_f=0$,
 and
${\tilde L}_0^{\rm int}-L_0^{\rm int}=\hf$. The simplest possibility
is an NS fermion from the current algebra of the internal theory;
however, such a state is eliminated by the GSO projection.
 Thus this model contains no
tachyons or massless states propagating in the black hole throat.

\bigskip\bigskip
\centerline{{\bf Acknowledgments}}\nobreak
I would like to thank
Steve Giddings, Andy Strominger, Joe Polchinski, Renata Kallosh
and Itzhak Bars for helpful
discussions, and
comments on earlier drafts.  This work was supported in
part by the grants DOE-91ER40618 and NSF PYI-9157463.

\listrefs

\end